\documentclass[aps,prd,twocolumn,groupedaddress]{revtex4}

\usepackage{enumitem}
\usepackage{graphicx}
\usepackage{dcolumn}
\usepackage{array,multirow}
\usepackage{bm}
\usepackage{amsmath}
\usepackage{epstopdf}
\usepackage{amsfonts}
\usepackage{amssymb}
\usepackage{epsfig}
\usepackage{tabularx}
\usepackage{color}
\usepackage[colorlinks=true,citecolor=blue,urlcolor=magenta,breaklinks]{hyperref}

\newcommand{\be}{\begin{equation}}
\newcommand{\en}{\end{equation}}
\newcommand{\bea}{\begin{eqnarray}}
\newcommand{\ena}{\end{eqnarray}}

\begin{document}

\title{Dark matter admixed strange quark stars in the Starobinsky model}

\author{Il\'idio Lopes and Grigoris Panotopoulos }
\email[]{grigorios.panotopoulos@tecnico.ulisboa.pt}
\email[]{ilidio.lopes@tecnico.ulisboa.pt}
\affiliation{Centro de Astrof{\'i}sica e Gravita\c c\~ao - CENTRA, Departamento de F{\'i}sica, Instituto Superior T{\'e}cnico - IST, Universidade de Lisboa - UL, Av. Rovisco Pais 1, 1049-001 Lisboa, Portugal
}

\date{\today}

\begin{abstract}
The properties of dark matter admixed strange quark stars are investigated in the Starobinsky model of modified gravity. For quark matter we assume the MIT bag model, while self-interacting dark matter inside the star is modelled as a Bose-Einstein condensate with a polytropic equation of state. We numerically integrate the structure equations in the Einstein frame adopting the two-fluid formalism treating the curvature correction term non-perturbatively. Our findings show that strange quark stars (in agreement with current observational constraints) with the highest masses are equally affected by dark matter and modified gravity.
\end{abstract}

%\keywords{Self-interacting dark matter \sep relativistic stars \sep alternative theories of gravity.}

\maketitle


\section{Introduction}

The $\Lambda CDM$ model, although very successful in describing a vast amount of observational data, it suffers from the cosmological constant problems \cite{weinberg}. Therefore, other possibilities have been explored, such as 
dynamical dark energy models \cite{dynamical} with a time varying equation-of-state parameter or modified gravity models, e.g. $f(R)$ theories of gravity \cite{modified1,modified2}, where the Ricci scalar $R$ in the Einstein-Hilbert term of General Relativity (GR) is replaced by a generic function. In addition, self-interacting dark matter has been proposed in order to alleviate some apparent conflicts between astrophysical observations and the collisionless dark matter paradigm \cite{self-interacting}.

\paragraph*{}

Nowadays the main motivation to study $f(R)$ theories of gravity is to explain the undergoing acceleration of the Universe, however one should investigate the astrophysical implications of this class of theories too. In compact stars \cite{textbook}, such as white dwarfs and neutron stars, matter is characterized by ultra high densities, and therefore they are excellent natural laboratories to study modified theories of gravity, since the extreme conditions they offer us cannot be reached in Earth-based experiments. Based on theoretical arguments a new class of compact objects has been postulated to exist -- strange quark stars.  There are some hints that such objects can exist in nature, and even explain the recent observations of super-luminous supernovae \cite{superluminous1,superluminous2}. This type of supernovae events occurs about one out of every 1000 supernovae explosions, and are more than 100 times brighter than normal supernovae. One plausible explanation for the formation of strange quark star is that neutrons are further compressed so that a new object made of de-confined quarks is formed \cite{strangestars1, strangestars2}. This new compact star  has a  much more stable matter configuration  than a regular neutron star. Its formation  could explain the origin of the huge amount of energy released in super-luminous supernovae. 

\paragraph*{}

Equally, it was shown recently that compact stars made entirely of self-interacting dark matter may also exist \cite{darkstars}.
In that work bosonic dark matter with a short-range repulsive potential was modelled inside the star as a Bose-Einstein condensate. In this scenario a polytropic equation-of-state of the form $P_\chi (\epsilon_\chi )=K_\chi \epsilon_\chi ^2$ was derived \cite{darkstars,chinos}, where the constant $K_\chi $ was found to be  $ K_\chi  = {2 \pi l_\chi }/{m_\chi^3}$,
with $m_\chi$ being the mass of the dark matter particle, while $l_\chi $ is the scattering length that determines the dark matter self-interaction cross section $\sigma_\chi = 4 \pi l_\chi ^2$. Although dark matter (DM) may not interact directly with ordinary matter, it can have significant gravitational effects.
Although the properties of dark matter inside stars have been studied in a large amount of literature  \cite[i.e.,][]{taoso,Martins2017,LKS2014,LS2014,LS2010,Kouvaris10,britoetal1,britoetal2}, the nature and origin of DM still remains a mystery, constituting one of the biggest challenges in modern theoretical cosmology.
The properties of  compact stars made of ordinary matter admixed with condensed dark matter have been studied in \cite{boson1,boson2,boson3,boson4}, and similarly admixed with fermionic matter in \cite{fermion1,fermion2,fermion3,fermion4,fermion5}. 

Although at first it was pointed out that $f(R)$ theories of gravity are unacceptable since they cannot support interior solutions of relativistic stars \cite{nosolution}, it was shown later that it is the properties of the trace of the matter energy-momentum tensor that determines whether relativistic stars exist or not \cite{langlois}. It is the goal of the present work to study condensed DM admixed strange quark stars in $f(R)$ theories of gravity, and in particular in the well motivated Starobinsky model $f(R)=R + aR^2$, i.e., a quadratic correction to the
Einstein-Hilbert term \cite{starobinsky}. Relativistic stars in $f(R)$ theories of gravity, and in particular in the Starobinsky model, have also been studied in \cite{fin1,fin2,greg,kokkotas,kokkotas1,kokkotas2,kokkotas3}.
 This study will help us to have some insight on how a strange DM star evolves in a $f(R)$ theory of gravity. During the process we will gain a better understating how DM interacts within a modified field of gravity.

Our work is organized as follows: In the next section we present the model as well as the observational constraints, while in section 3 we present the structure equations, which are integrated numerically, and we discuss our numerical results. We finish concluding our work in the last section.

\section{Theoretical framework}

\subsection{Alternative theory of gravity}

The model in the so-called Jordan frame is described by the action
\begin{equation}
S = \frac{1}{16 \pi G} \int d^4x \sqrt{-g} f(R) + S_M[\psi_i, g_{\mu \nu}]
\end{equation}
where $G$ is Newton's constant, $g_{\mu \nu}$ is the metric tensor, $R$ is the Ricci scalar, and $S_M$ is the matter
action that depends on the metric tensor and the matter fields $\psi_i$. 
In the following we shall be considering the Starobinsky model $f(R)=R+aR^2$ \cite{starobinsky}, where $a$ is the only free parameter of the gravitational theory, and clearly the $a=0$ case corresponds to GR. The parameter $a$ has dimensions $[mass]^{-2}$, and therefore it can be written also in the form $a=1/M^2$, where now the mass scale $M$ is the free parameter of the theory.
From a theoretical point of view the R-squared gravity is well-motivated, since higher order in $R$ terms are natural in Lovelock theory \cite{lovelock}, and also higher order curvature corrections appear in the low-energy effective equations of Superstring Theory \cite{ramgoolam}.
 
As usual we choose to work in the Einstein frame by performing a conformal transformation \cite{kokkotas,kokkotas1,kokkotas2,woodard,davis}
\begin{equation}
\tilde{g}_{\mu \nu} 
= A^{-2} g_{\mu \nu}
\end{equation}
where $A=e^{-\phi/\sqrt{3}}$, and where the action takes the equivalent form \cite{kokkotas,kokkotas1,kokkotas2,kokkotas3,woodard,davis}
\begin{widetext}
	\begin{equation}
	S = \frac{1}{16 \pi G} \int d^4x \sqrt{-\tilde{g}} [\tilde{R}-2 \tilde{g}^{\mu \nu} \partial_\mu \phi \partial_\nu \phi-V] + S_M[\psi_i, \tilde{g}_{\mu \nu} A^2]
	\end{equation}
\end{widetext}
Clearly the system looks like GR with an extra scalar field with a self-interaction potential $V$ given by \cite{kokkotas1,kokkotas2,woodard}
\begin{equation}
V\equiv V(\phi)
= \frac{(1-e^{-2 \phi/\sqrt{3}})^2}{4 a}.
\end{equation}
Expanding in powers of $\phi$ it is easy to see that the mass of the scalar field is given by $ m_\phi^2 = {1}/{(6 a)}= {M^2}/{6} $, and therefore to avoid tachyonic instabilities $m_\phi^2 < 0$, we require $a > 0$. 

\paragraph*{}

Varying with respect to the metric tensor and the scalar field one obtains Einstein's field equations as well as the Klein-Gordon equation
\begin{eqnarray}
\tilde{G}_{\mu \nu} & = & 8 \pi G [ \tilde{T}_{\mu \nu} + T^{\phi}_{\mu \nu} ] \\
\nabla_\mu \nabla^\mu{\phi}-\frac{1}{4} V_{,\phi} & = & -4 \pi G \alpha \tilde{T}
\end{eqnarray}
where $T^{\phi}_{\mu \nu}$ is the stress-energy tensor corresponding to the scalar field, "$,\phi$" denotes differentiation with
respect to the scalar field, while due to the conformal transformation there is a direct coupling between matter and the scalar field with the coupling constant being $\alpha=-1/\sqrt{3}$ \cite{kokkotas,kokkotas1,kokkotas2,kokkotas3}.
The matter energy-momentum tensor $T_{\mu \nu}$ in the Jordan frame and $\tilde{T}_{\mu \nu}$ in the Einstein frame are 
related via \cite{kokkotas} $ \tilde{T}_{\mu \nu} = A^2 T_{\mu \nu} $,
and in particular in the case of a perfect fluid the energy densities and the pressures in the two frames are related 
via $ \tilde{\epsilon} =  A^4 \epsilon $ and  \cite{kokkotas} $\tilde{P}  =  A^4 P $, where the tilde indicates the Einstein frame.

\subsection{Equations of state}

For ordinary quark matter we shall consider the simplest equation of state corresponding to a relativistic gas of de-confined quarks,
known also as the MIT bag model \cite{MIT1,MIT2,MIT3}
\begin{equation}
P_s = \frac{1}{3} (\epsilon_s - 4B)
\end{equation}
and the bag constant has been taken to be $B^{1/4}=148 MeV$ \cite{Bvalue}.
Although refinements of the bag model exist in
the literature \cite{refine1,refine2,refine3} (for the present state-of-the-art 
see the recent paper \cite{art}), the above analytical expression "radiation plus constant" has been employed in recent works, both in GR \cite{prototype}, where it was shown that the observed value of the cosmological constant $\Lambda \sim (10^{-33} eV)^2$ is too small to have an effect, and in R-squared gravity \cite{kokkotas1,kokkotas3}. Therefore, despite its simplicity, we will consider the MIT bag model in the present work, since it suffices for our purposes, and we will take the cosmological constant to be zero.

\paragraph*{}

For the condensed DM we shall consider the equation of state obtained in \cite{darkstars}, namely $P_\chi=K_\chi \epsilon_\chi^2$,
where the constant $K_\chi=2 \pi l_\chi/m_\chi^3$ is given in terms of the mass of the dark matter particles $m_\chi$ and the scattering length $l_\chi$.
In a dilute and cold gas only the binary collisions at low energy are relevant, and these collisions are characterized by the s-wave scattering length
$l_\chi$ independently of the form of the two-body potential \cite{darkstars}. Therefore we can consider a short range repulsive delta-potential of the form
$V(\vec{r}_1-\vec{r}_2) = ({4 \pi l_\chi}/{m_\chi}) \delta^{(3)}(\vec{r}_1-\vec{r}_2)
$, which implies a dark matter self interaction cross section of the form $\sigma_\chi=4 \pi l_\chi^2$ \cite{darkstars,chinos}.

\subsection{Observational constraints}

Since a light scalar field can mediate a long range attractive force, one expects modifications to the Newtonian potential $1/r$. Therefore, the mass of the scalar field must be sufficiently high in order to be compatible with solar system tests. The expression of the Post-Newtonian parameter $\gamma$, which has been measured by the Cassini mission, in $f(R)$ theories of gravity was obtained in \cite{postnewtonian}. On the one hand the parameter $\gamma$ in $f(R)$ theories it has been computed to be
$ \gamma = ({3-e^{-Mr/\sqrt{6}}})/({3+e^{-Mr /\sqrt{6}}})$,
where the function $e^{-Mr/\sqrt{6}}$ corresponds to
$e^{-m_\phi r}$ with $m_\phi$ being the mass of the scalar field. For the Starobinsky model $m_\phi=M/\sqrt{6}$. On the other hand the measurement according to the Cassini mission is the following \cite{cassini}
$ \gamma = 1 + (2.1 \pm 2.3) \times 10^{-5} $ for which $r=1.5 \times 10^8 km$. Therefore the mass scale $M$ characterizing the Starobinsky model must satisfy the condition $ M \geq 4 \times 10^{-26} GeV $.

\paragraph*{}

What is more, current observations constrain self-interacting dark matter \cite{self-interacting,bullet1,bullet2,review}, and here we apply the bounds $0.45\, {cm^2}{g^{-1}} < {\sigma_\chi}/{m_\chi} < 1.5\, {cm^2}{g^{-1}}$. By fixing $l_\chi=1 fm$ the above condition implies the following range for the mass of the dark matter particle $ 0.05\, GeV < m_\chi < 0.16\, GeV $, and thus the lowest and highest values of $K_\chi$ are
determined by the relation $K_\chi=2\pi l_\chi/m_\chi^3$.

\smallskip

\paragraph*{}
Therefore in the following we shall consider: 
\begin{enumerate}[label=(\alph*)]
\item	
for the $a$ parameter in Starobinsky's model several values in the range
\begin{equation}
\frac{5 \times 10^{76}}{m_{pl}^2} \leq a \leq \frac{10^{88}}{m_{pl}^2},
\end{equation}
where $m_{pl}=1.22 \times 10^{22}~GeV$ is the Planck mass.

\item	
regarding the DM equation-of-state $K_\chi$ values that respect the limits on the self-interaction cross section, i.e., 
\begin{equation}
\frac{4}{B} < K_\chi < \frac{150}{B},
\end{equation}
where now the constant $K_\chi$ is given in units of the bag constant.
\end{enumerate}
The values of the $a$ parameter considered here are also compatible with
observations regarding the emission of gravitational radiation from binary systems. The calculation in the framework of GR can be found in \cite{GR1,GR2}, while for a similar calculation in the framework of $f(R)$ theories of gravity see \cite{Laurentis1,Laurentis2,Laurentis3}.

\medskip

\section{Numerical results}

Seeking static spherically symmetric solutions of the form
\begin{eqnarray}
ds^2 = -e^{2 \nu(r)} dt^2 + e^{2 \lambda(r)} dr^2 + r^2 d\Omega^2,
\label{eq:ds2}
\end{eqnarray}
where $d\Omega^2\equiv d \theta^2 + sin^2 \theta d \varphi^2$.
\smallskip
one obtains the following structure equations \cite{kokkotas}
\begin{widetext}
	\begin{eqnarray}
	\frac{1}{r^2} \frac{d}{d r} [r (1-e^{-2 \lambda} )] & = & 8 \pi G A^4 \epsilon + \frac{V}{2} + e^{-2 \lambda} \left( \frac{d \phi}{d r} \right)^2 \\
	\frac{2}{r}e^{-2 \lambda}\frac{d \nu}{d r}-\frac{1}{r^2} (1-e^{-2 \lambda} )) & = & 8 \pi G A^4 P - \frac{V}{2} + e^{-2 \lambda}\left( \frac{d \phi}{d r} \right)^2 \\
	\frac{d^2 \phi}{dr^2} + \left( \frac{2}{r}+\frac{d \nu}{d r}-\frac{d \lambda}{d r} \right) \frac{d \phi}{d r} & = & 4 \pi G \alpha A^4 (\epsilon-3 P) e^{2 \lambda} + \frac{1}{4} V_{,\phi} e^{2 \lambda}\\
	\frac{d P}{d r} & = & - (P+\epsilon) \left( \frac{d \nu}{d r} + \alpha \frac{d \phi}{d r} \right)
	\end{eqnarray}
\end{widetext}
where $P$ and $\epsilon$ are the pressure and the energy density respectively of a single fluid. Clearly when the scalar field is absent one recovers the usual
Tolman-Oppenheimer-Volkoff equations \cite{TOV1,TOV2}. 

The system of coupled equations is supplemented with  the initial conditions at the center of the star ($r=0$):
$ \epsilon(0) = \epsilon_c $, $ \lambda(0) = 0 $
$\phi(0)  = \phi_c $ and $ {d \phi}/{d r}(0) = 0$,
where $\epsilon_c$ and $\phi_c$ are the central values of the DM energy density and of the scalar field respectively, and the last condition ensures the regularity of the scalar field. In principle one could handle the problem in perturbations theory, assuming that the parameter $a$ is small and therefore the higher order term in $R$ just perturbs the GR solution. However, it was shown in \cite{kokkotas} that in dealing with relativistic stars an analysis based on perturbation theory does not provide us with reliable results, and therefore in the present work we shall treat the problem exactly, i.e. non-perturbatively.
We wish to stress the fact that contrary to $\epsilon_c$, $\phi_c$ is not an arbitrary initial value but it must be determined self-consistently. 
In the exterior problem solution all three quantities $P=\epsilon=\phi=0$ are zero, and therefore when we consider the interior problem solution we must require that both the dark matter energy density and the scalar field vanish on the surface of the star radius ($r=R_*$):
 $ \epsilon(R_*)  =  0 $ and $ \phi(R_*)  =  0 $,
which clearly is not possible for any $\phi_c$. Therefore fixing the numerical values of the parameters $a, K, \epsilon_c$, the scalar field central value $\phi_c$ is determined requiring that $\phi$ and $\epsilon$ vanish simultaneously at the surface of the star. As usual the radius of the star $R_*$ is determined by the condition $P(R_*)=0$, while the mass of the star $M_*$ is given by $M_*=m(R_*)$, where the new function $m(r)$ is defined as follows
\begin{equation}
1-\frac{2 m(r)}{r}=e^{-2 \lambda(r)}.
\end{equation}
Finally, if inside the star there two fluids without direct interaction, we need to work in the two-fluid formalism \cite{2fluid1,2fluid2}, according to which in the first 3 equation above $P$ ($\equiv P_s+P_\chi$) and $\epsilon $
($\equiv \epsilon_s+\epsilon_\chi$) are the total pressure
and the total energy density respectively, while there is a set of
equations:
\begin{equation}
\frac{d P_i}{d r} = - (P_i+\epsilon_i) \left( \frac{d \nu}{d r} + \alpha \frac{d \phi}{d r} \right),
\end{equation}
one for each fluid separately ($i=s,\chi$).

\begin{figure}
\centering
\includegraphics[scale=0.8]{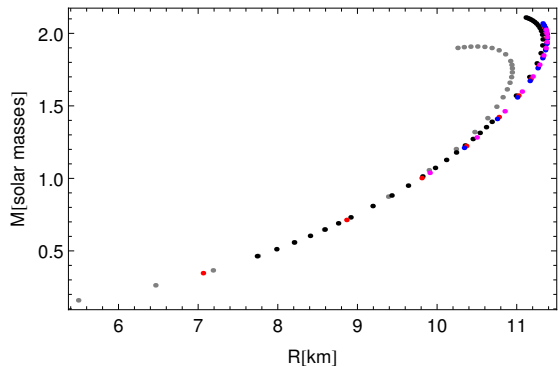}
\caption{Mass-to-radius profile for DM admixed strange quark star
(with $K_\chi=4/B$ and $f_\chi=0.1$) in a f(R) theory of gravity with 3 different values of the $a$ parameter: $a=10^{77}$ (red curve), $a=10^{78}$ (blue curve), $a=10^{79}$ (magenta curve). The profiles of a strange quark star in GR ($a=0$) without DM (gray curve) and with DM (black curve) are also shown for comparison.}
\label{fig:MR} 	
\end{figure}

\begin{figure}
\centering
\includegraphics[scale=0.8]{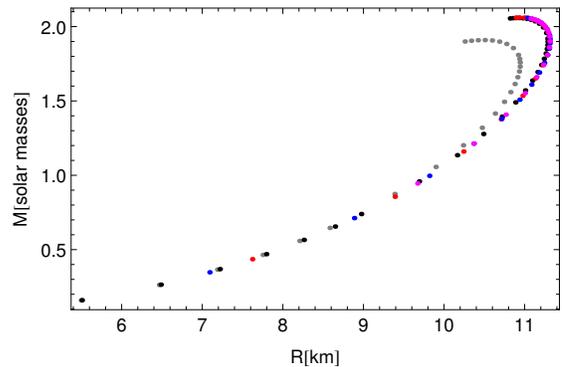}
\caption{Mass-to-radius profile for DM admixed strange quark star
(with $K_\chi=150/B$ and $f_\chi=0.05$) in a f(R) theory of gravity with 3 different values of the $a$ parameter: $a=5 \times 10^{76}$ (red curve), $a=10^{80}$ (blue curve), $a=10^{88}$ (magenta curve).}	
\label{fig:MR2} 	
\end{figure}

\paragraph*{}

In this case in order to integrate the structure equations we need to specify the central values both for ordinary quark matter and for dark matter $P_s(0)$ and $P_{\chi}(0)$ respectively. So in the following a certain DM scenario is defined by both the constant $K_\chi=2 \pi l_\chi/m_\chi^3$ and the dark matter fraction
\begin{equation}
f_\chi = \frac{P_{\chi}(0)}{P_s(0)+P_{\chi}(0)}.
\end{equation}

Figures 1-6 summarize our numerical results. Figure~\ref{fig:MR} shows the mass-to-radio profile of a DM admixed strange quark star
for the DM scenario $K_\chi=4/B$, 
$ f_\chi=0.1$ and 3 different values of $a=10^{77},10^{78},10^{79}$ (in units where $G=1=m_{pl}$).
Figure~\ref{fig:MR2} is similar to Figure~\ref{fig:MR}, it shows the mass-to-radio profile now for the DM scenario $K_\chi=150/B$, $f_\chi=0.05$ and for $f(R)$ theory of gravity with the values of $a=5 \times 10^{76}, 10^{80}, 10^{88} $.

\begin{figure}
\centering
\includegraphics[scale=0.8]{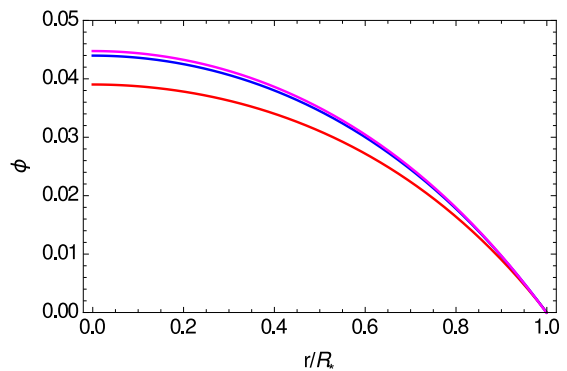}
\caption{Scalar field $\phi$ as a function of the dimensionless star radius $r/R_*$. The compact star is an admixed DM strange quark star for which $K_\chi=150/B$ and $f_\chi=0.05$. The scalar field is shown for 3 different values of $a$: $a=5 \times 10^{76}$ (red), $a=3 \times 10^{77}$ (blue), $a=10^{78}$ (magenta) and central total pressure $P(0)=B$ (or $0.36 \%$ of DM).}
\label{fig:scalaron} 	
\end{figure}

\begin{figure}
\centering
\includegraphics[scale=0.8]{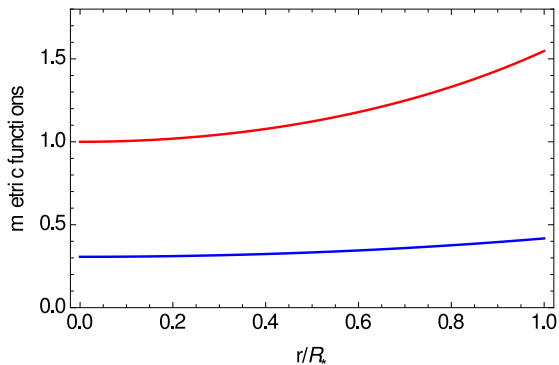}
\caption{Same as figure~\ref{fig:scalaron} but for the two metric functions (equation~\ref{eq:ds2}): $e^{2\nu}$ (blue curve) and $e^{2\lambda}$ (red curve).
The metric functions coincide for all 3 values of $a$ (see figure~\ref{fig:scalaron}), so there is only one set of curves.}
	\label{fig:metric} 	
\end{figure}

\begin{figure}
\centering
\includegraphics[scale=0.8]{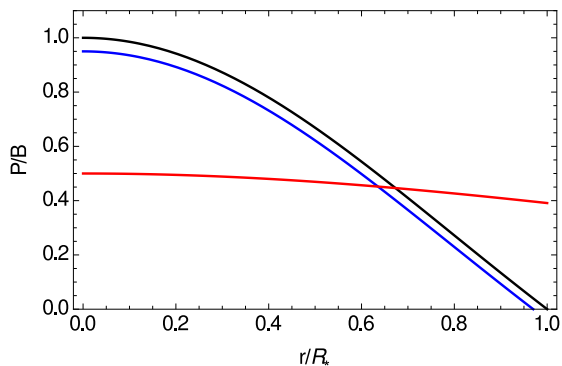}
\caption{Same as figure~\ref{fig:scalaron} but for the normalized pressure $P_i/B$ ($i=s,\chi$)  of the fluids. Shown are the curves of the total pressure in black,the  quark pressure in blue and the DM pressure in red. The metric functions coincide for all 3 values of $a$ (see figure~\ref{fig:scalaron}), so there is only one set of curves. Note that curve corresponding to DM pressure (red curve) has been scaled to 10:1 for better visualization.}
\label{fig:pressure} 	
\end{figure}

Figure~\ref{fig:scalaron} shows the scalar field $\phi$ versus the star's dimensionless radius $r/R_*$, while figure~\ref{fig:metric} and~\ref{fig:pressure} show the metric functions and the normalized pressure $P/B$ of the fluids, respectively, as a function of the star's dimensionless radius $r/R_*$ for the same DM scenario ($K_\chi=150/B$ and $f_\chi=0.05$), central total pressure $P(0)=B$, and for $a=5 \times 10^{76}, 3 \times 10^{77}, 10^{78}$ (or $0.36 \%$ of DM). Varying $a$ does not have any effect on the pressure and metric (cf. Figures~\ref{fig:pressure} and~\ref{fig:metric}), but it has some impact on the solution for the scalar field (cf. Figure~\ref{fig:scalaron}). Finally, the last figure shows the mass-to-radius profiles for two DM models, namely model $A$ with $K_\chi=4/B$ and $f_\chi=0.09$, and model $B$ with $K_\chi=150/B$ and  $f_\chi=0.05$, both in GR ($a=0$) and in Starobinsky's model for $a=10^{77})$. In all 3 figures showing the mass-to radius profiles, the difference compared to the standard curve corresponding to GR without DM appears in the upper part of the curves for star masses higher than at least one solar mass.

Our numerical results indicate that alternative $f(R)-$ theory of gravity or the presence of DM inside the star predict very similar mass-to-radius profiles.
Bosonic DM, however, can give potential signatures that can never occur in modified theories of gravity. For example, in particle physics models with a rich Higgs sector with both neutral and charged Higgs bosons, two DM particles can annihilate to a pair of photons via loop Feynman diagrams in which charged Higgs bosons circulate, see e.g. \cite{Wang:2012ts}. These monochromatic gamma rays could be detected by the Fermi telescope \cite{LAT}.

\begin{figure}
$$\qquad$$	
\centering
\includegraphics[scale=0.8]{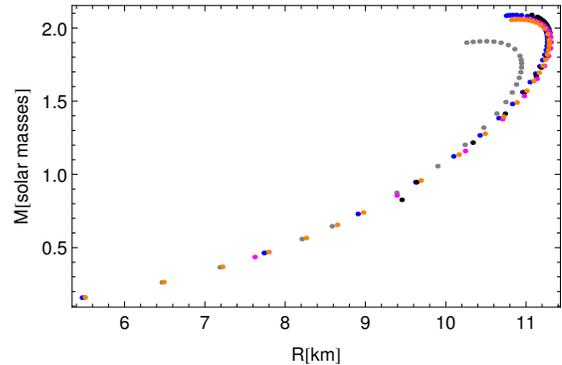}
\caption{Mass-to-radius profiles for two DM admixed strange quark star in an f(R) theory of gravity (with  $a=10^{77})$: model $A$ (black curve) with $K_\chi=4/B$ and $f_\chi=0.09$; and model $B$ (magenta curve) with $K_\chi=150/B$ and  $f_\chi=0.05$. For comparison, we also show the same models of DM admixed strange quark stars in GR ($a=0$), model A (blue curve) and model B (orange curve). The same strange quark star without DM in GR is also shown for comparison(gray curve).}\label{fig:2DM} 	
\end{figure}

\section{Conclusions}

The properties of DM admixed strange quark stars in the Starobinsky model $f(R)=R+aR^2$ have been investigated, and the structure equations have benn integrated numerically to produce the mass-to-radius profiles for the relativistic compact stars. Dark matter is assumed to be bosonic and self-interacting characterized by a DM particle mass $m_\chi$ and a scattering length $l_\chi$, and it is modelled inside the star as a Bose-Einstein condensate leading to a polytropic EoS. 
We have assumed that ordinary quark matter and DM interact only gravitationally, we have adopted the two-fluid formalism in the Einstein frame, and we have treated the higher order in $R$ term non-perturbatively.
Considering values of $a, m_\chi, l_\chi$ compatible with current data, we have investigated the effect on the properties of the stars, such as total pressure, scalar field solution and mass-to-radius profiles, of the amount of DM as well as the higher order in $R$ term. We found that for strange stars the impact of condensed admixed bosonic dark matter and $f(R)$ modified gravity affect equally the higher masses of compact stars. This result shows that known observational effects of gravity observed in the Universe could be equally attributed to a modified gravity or dark matter particles.

%%%%%%%%%%%%%%%%%%%%%%%%%%%%%%%%%%%%%%%%%%%%%%%%%%%%%%%%%%%%%%%%%%%%%%%%%%%%%%%%%%%%%%%%%%%%%%%%%%%%%%%%%%%%%%%%%%%%%%%%%%%%%%%%%%%%

\section*{Acknowlegements}

The authors thank the Funda\c c\~ao para a Ci\^encia e Tecnologia (FCT), Portugal, for the financial support to the Center for Astrophysics and Gravitation-CENTRA, Instituto Superior T\'ecnico, Universidade de Lisboa,  through the Grant No. UID/FIS/00099/2013.

%%%%%%%%%%%%%%%%%%%%%%%%%%%%%%%%%%%%%%%%%%%%%%%%%%%%%%%%%%%%%%%%%%%%%%%%%%%

%%%%%%%%%%%%%%%%%%%%%%%%%%%%%%%%%%%%%%%%%%%%%%%%%%%%%%%%%%%%%%%%%%%%%%%%%%%%%%%%%%%%%%%%%%%%%%%%%%%%%%%%%%%%%%%%%%%%%%%%%%%%%%%%%%%%


\begin{thebibliography}{99}
	\bibitem{weinberg} S.~Weinberg,
	%``The Cosmological Constant Problem,''
	Rev.\ Mod.\ Phys.\  {\bf 61} (1989) 1.
	
	\bibitem{dynamical} E.~J.~Copeland, M.~Sami and S.~Tsujikawa,
	%``Dynamics of dark energy,''
	Int.\ J.\ Mod.\ Phys.\ D {\bf 15} (2006) 1753
	[hep-th/0603057].
	
	\bibitem{modified1} T.~P.~Sotiriou and V.~Faraoni,
	%``f(R) Theories Of Gravity,''
	Rev.\ Mod.\ Phys.\  {\bf 82} (2010) 451
	[arXiv:0805.1726 [gr-qc]].
	
	\bibitem{modified2} A.~De Felice and S.~Tsujikawa,
	%``f(R) theories,''
	Living Rev.\ Rel.\  {\bf 13} (2010) 3
	[arXiv:1002.4928 [gr-qc]].
	
	\bibitem{self-interacting} D.~N.~Spergel and P.~J.~Steinhardt,
	%``Observational evidence for selfinteracting cold dark matter,''
	Phys.\ Rev.\ Lett.\  {\bf 84} (2000) 3760
	[astro-ph/9909386].
	
	%\bibitem[Casanellas et al.(2012)]{2012ApJ...745...15C} J.~Casanellas, P.~Pani, I.~Lopes, V.~ Cardoso,
	%Testing Alternative Theories of Gravity Using the Sun.\ 
	%The Astrophysical Journal {\bf 745}, 15 [arXiv:1109.0249  [astro-ph]]. 
	
	%\bibitem[Pani et al.(2011)]{2011PhRvL.107c1101P} P.~ Pani, V.~Cardoso, T.~Delsate, 
	%Compact Stars in Eddington Inspired Gravity.\ 
	%Physical Review Letters {\bf 107} (2011), 031101
	%[arXiv:1106.3569 [gr-qc]].
	
	%\bibitem{britoetal1} R.~Brito, V.~ Cardoso, H.~Okawa,
	%``ccretion of Dark Matter by Stars,''
	%Phys.\ Rev.\ Lett.\  {\bf 115} (2015) 111301
	%[arXiv:1508.04773].
	
	%\bibitem{britoetal2} R.~Brito, V.~ Cardoso, H.~Okawa,
	%R.~Brito, V.~Cardoso, C.~F.~B.~Macedo, H.~Okawa, C.~Palenzuela,
	%``Interaction between bosonic dark matter and stars,''
	%Phys.\ Rev.\ D {\bf 93} (2016)  044045
	%[arXiv:1512.00466].
	
	%\bibitem{LS2010} Lopes, I., Silk, J.
	%\ 2010.\ Neutrino Spectroscopy Can Probe the Dark Matter Content in the Sun.\ 
	%Science {\bf 330} (2010), 462. 
	
	%\bibitem{Kouvaris10} C. Kouvaris, P. Tinyakov, 
	%\ Can neutron stars constrain dark matter?.\  
	%Phys.\ Rev.\ D {\bf 82} (2010)  063531
	%[arXiv:1004.0586].
	
	%\bibitem{LS2012} I. Lopes , J. Silk, \
	%\ Solar Constraints on Asymmetric Dark Matter.
	%The Astrophysical Journal {\bf 757}  (2012), 130 	arXiv:1209.3631 [astro-ph.SR]].
	
	\bibitem{textbook} S.~L.~Shapiro and S.~A.~Teukolsky,
	``Black holes, white dwarfs, and neutron stars: The physics of compact objects,''
	New York, USA: Wiley (1983) 645 p.
	
	\bibitem{superluminous1} E.~O.~Ofek {\it et al.},
	%``SN 2006gy: An extremely luminous supernova in the early-type galaxy NGC 1260,''
	Astrophys.\ J.\  {\bf 659} (2007) L13
	[astro-ph/0612408].
	
	\bibitem{superluminous2} R.~Ouyed, D.~Leahy and P.~Jaikumar,
	%``Predictions for signatures of the quark-nova in superluminous supernovae,''
	arXiv:0911.5424 [astro-ph.HE].
	
	\bibitem{strangestars1} Y.~L.~Yue, X.~H.~Cui and R.~X.~Xu,
	%``Is psr b0943+10 a low-mass quark star?,''
	Astrophys.\ J.\  {\bf 649} (2006) L95
	[astro-ph/0603468].
	
	\bibitem{strangestars2} D.~Leahy and R.~Ouyed,
	%``Supernova SN2006gy as a first ever Quark Nova?,''
	Mon.\ Not.\ Roy.\ Astron.\ Soc.\  {\bf 387} (2008) 1193
	[arXiv:0708.1787 [astro-ph]].
	
	\bibitem{darkstars} X.~Y.~Li, T.~Harko and K.~S.~Cheng,
	%``Condensate dark matter stars,''
	JCAP {\bf 1206} (2012) 001
	[arXiv:1205.2932 [astro-ph.CO]].
	
	\bibitem{chinos} X.~Li, F.~Wang and K.~S.~Cheng,
	%``Gravitational effects of condensate dark matter on compact stellar objects,''
	JCAP {\bf 1210} (2012) 031
	[arXiv:1210.1748 [astro-ph.CO]].
	
	
	\bibitem{taoso} M.~Taoso, G.~Bertone and A.~Masiero,
	%``Dark Matter Candidates: A Ten-Point Test,''
	JCAP {\bf 0803} (2008) 022
	[arXiv:0711.4996 [astro-ph]].
	
	
	\bibitem{LS2014} I. Lopes , J. Silk, \
	%\ A Particle Dark Matter Footprint on the First Generation of Stars.\
	The Astrophysical Journal {\bf 786}  (2014), 25 [arXiv:1404.3909]
	
	
	\bibitem{LS2010} I. Lopes , J. Silk, \
	%\ Probing the Existence of a Dark Matter Isothermal Core Using Gravity Modes.\
	The Astrophysical Journal {\bf 722} (2010), L95-L99 [arXiv:1009.5122]
	
	% IPL
	\bibitem{Martins2017} A/ Martins, A., I. Lopes, J. Casanellas, 
	%\ Asteroseismic constraints on asymmetric dark matter: Light particles with an effective spin-dependent coupling.
	Phys.\ Rev.\ D {\bf 95} (2017)  023507
	[arXiv:1701.03928].
	
	\bibitem{LKS2014} I. Lopes,Kadota, K., , J. Silk, \
	%\ Constraint on Light Dipole Dark Matter from Helioseismology.\
	The Astrophysical Journal Letters {\bf 780}  (2014), L15 [arXiv:1310.0673]
	
	\bibitem{Kouvaris10} C. Kouvaris, P. Tinyakov, 
	%\ Can neutron stars constrain dark matter?.\  
	Phys.\ Rev.\ D {\bf 82} (2010)  063531
	[arXiv:1004.0586].
	
	\bibitem{britoetal1} R.~Brito, V.~ Cardoso, H.~Okawa,
	%``ccretion of Dark Matter by Stars,''
	Phys.\ Rev.\ Lett.\  {\bf 115} (2015) 111301
	[arXiv:1508.04773].
	
	\bibitem{britoetal2} R.~Brito, V.~ Cardoso, H.~Okawa,
	R.~Brito, V.~Cardoso, C.~F.~B.~Macedo, H.~Okawa, C.~Palenzuela,
	%``Interaction between bosonic dark matter and stars,''
	Phys.\ Rev.\ D {\bf 93} (2016)  044045
	[arXiv:1512.00466].
	
	\bibitem{boson1} X.~Li, F.~Wang and K.~S.~Cheng,
	%``Gravitational effects of condensate dark matter on compact stellar objects,''
	JCAP {\bf 1210} (2012) 031
	[arXiv:1210.1748 [astro-ph.CO]].
	
	\bibitem{boson2} A.~Li, F.~Huang and R.~X.~Xu,
	%``Too massive neutron stars: The role of dark matter?,''
	Astropart.\ Phys.\  {\bf 37} (2012) 70
	[arXiv:1208.3722 [astro-ph.SR]].
	
	\bibitem{boson3} G.~Panotopoulos and I.~Lopes,
	%``Gravitational effects of condensed dark matter on strange stars,''
	Phys.\ Rev.\ D {\bf 96} (2017) no.2,  023002
	[arXiv:1706.07272 [gr-qc]].
	
	\bibitem{boson4} G.~Panotopoulos and I.~Lopes,
	%``Radial oscillations of strange quark stars admixed with condensed dark matter,''
	Phys.\ Rev.\ D {\bf 96} (2017) no.8,  083013
	[arXiv:1709.06643 [gr-qc]].
	
	\bibitem{fermion1} G.~Narain, J.~Schaffner-Bielich and I.~N.~Mishustin,
	%``Compact stars made of fermionic dark matter,''
	Phys.\ Rev.\ D {\bf 74} (2006) 063003
	[astro-ph/0605724].
	
	\bibitem{fermion2} S.~C.~Leung, M.~C.~Chu and L.~M.~Lin,
	%``Dark-matter admixed neutron stars,''
	Phys.\ Rev.\ D {\bf 84} (2011) 107301
	[arXiv:1111.1787 [astro-ph.CO]].
	
	\bibitem{fermion3} S.~C.~Leung, M.~C.~Chu and L.~M.~Lin,
	%``Equilibrium Structure and Radial Oscillations of Dark Matter Admixed Neutron Stars,''
	Phys.\ Rev.\ D {\bf 85} (2012) 103528
	[arXiv:1205.1909 [astro-ph.CO]].
	
	\bibitem{fermion4} P.~Mukhopadhyay and J.~Schaffner-Bielich,
	%``Quark stars admixed with dark matter,''
	Phys.\ Rev.\ D {\bf 93} (2016) no.8,  083009
	[arXiv:1511.00238 [astro-ph.HE]].
	
	\bibitem{fermion5} G.~Panotopoulos and I.~Lopes,
	%``Dark matter effect on realistic equation of state in neutron stars,''
	Phys.\ Rev.\ D {\bf 96} (2017) no.8,  083004
	[arXiv:1709.06312 [hep-ph]].
	
	\bibitem{nosolution} T.~Kobayashi and K.~i.~Maeda,
	%``Relativistic stars in f(R) gravity, and absence thereof,''
	Phys.\ Rev.\ D {\bf 78} (2008) 064019
	[arXiv:0807.2503 [astro-ph]].
	
	\bibitem{langlois} E.~Babichev and D.~Langlois,
	%``Relativistic stars in f(R) gravity,''
	Phys.\ Rev.\ D {\bf 80} (2009) 121501
	Erratum: [Phys.\ Rev.\ D {\bf 81} (2010) 069901]
	[arXiv:0904.1382 [gr-qc]].
	
	\bibitem{starobinsky} A. A. Starobinsky, Phys. Lett. B 91 (1980) 99.
	
	\bibitem{fin1} K.~Kainulainen, V.~Reijonen and D.~Sunhede,
	%``The Interior spacetimes of stars in Palatini f(R) gravity,''
	Phys.\ Rev.\ D {\bf 76} (2007) 043503
	[gr-qc/0611132].
	
	\bibitem{fin2} K.~Kainulainen, J.~Piilonen, V.~Reijonen and D.~Sunhede,
	%``Spherically symmetric spacetimes in f(R) gravity theories,''
	Phys.\ Rev.\ D {\bf 76} (2007) 024020
	[arXiv:0704.2729 [gr-qc]].
	
	\bibitem{greg} G.~Panotopoulos,
	%``Strange stars in $f(R)$ theories of gravity in the Palatini formalism,''
	Gen.\ Rel.\ Grav.\  {\bf 49} (2017) no.5,  69
	[arXiv:1704.04961 [gr-qc]].
	
	\bibitem{kokkotas} S.~S.~Yazadjiev, D.~D.~Doneva, K.~D.~Kokkotas and K.~V.~Staykov,
	%``Non-perturbative and self-consistent models of neutron stars in R-squared gravity,''
	JCAP {\bf 1406} (2014) 003
	[arXiv:1402.4469 [gr-qc]].
	
	\bibitem{kokkotas1} K.~V.~Staykov, D.~D.~Doneva, S.~S.~Yazadjiev and K.~D.~Kokkotas,
	%``Slowly rotating neutron and strange stars in $R^2$ gravity,''
	JCAP {\bf 1410} (2014) no.10,  006
	[arXiv:1407.2180 [gr-qc]].
	
	\bibitem{kokkotas2} S.~S.~Yazadjiev, D.~D.~Doneva and K.~D.~Kokkotas,
	%``Rapidly rotating neutron stars in R-squared gravity,''
	Phys.\ Rev.\ D {\bf 91} (2015) no.8,  084018
	[arXiv:1501.04591 [gr-qc]].
	
	\bibitem{kokkotas3} K.~V.~Staykov, D.~D.~Doneva, S.~S.~Yazadjiev and K.~D.~Kokkotas,
	%``Gravitational wave asteroseismology of neutron and strange stars in R$^2$ gravity,''
	Phys.\ Rev.\ D {\bf 92} (2015) no.4,  043009
	[arXiv:1503.04711 [gr-qc]].
	
	\bibitem{lovelock} D.~Lovelock,
	%``The Einstein tensor and its generalizations,''
	J.\ Math.\ Phys.\  {\bf 12} (1971) 498.
	
	\bibitem{ramgoolam} S.~Corley, D.~A.~Lowe and S.~Ramgoolam,
	%``Einstein-Hilbert action on the brane for the bulk graviton,''
	JHEP {\bf 0107} (2001) 030
	[hep-th/0106067].
	
	\bibitem{woodard} R.~P.~Woodard,
	%``Avoiding dark energy with 1/r modifications of gravity,''
	Lect.\ Notes Phys.\  {\bf 720} (2007) 403
	[astro-ph/0601672].
	
	\bibitem{davis} P.~Brax, C.~van de Bruck, A.~C.~Davis and D.~J.~Shaw,
	%``f(R) Gravity and Chameleon Theories,''
	Phys.\ Rev.\ D {\bf 78} (2008) 104021
	[arXiv:0806.3415 [astro-ph]].
	
	\bibitem{MIT1} Chodos A, Jaffe R L, Johnson K, Thorn C B, and Weisskopf V F,
	1974a Phys. Rev. D. 9 3471-95.
	
	\bibitem{MIT2} Chodos A, Jaffe R L, Johnson K, and Thorn C B,
	1974b Phys. Rev. D. 10 2599-2604.
	
	\bibitem{MIT3} Farhi E and Jaffe R L 1984 Phys. Rev. D. 30 2379.
	
	\bibitem{Bvalue} P. Haensel, J. L. Zdunik and R. Schaeffer,  Astron. Astrophys. {\bf 160} (1986) 121.
	
	\bibitem{refine1} V.~D.~Toneev, E.~G.~Nikonov, B.~Friman, W.~Norenberg and K.~Redlich,
	%``Strangeness production in nuclear matter and expansion dynamics,''
	Eur.\ Phys.\ J.\ C {\bf 32} (2003) 399
	[hep-ph/0308088].
	
	\bibitem{refine2} Y.~B.~Ivanov, A.~S.~Khvorostukhin, E.~E.~Kolomeitsev, V.~V.~Skokov, V.~D.~Toneev and D.~N.~Voskresensky,
	%``Lattice QCD constraints on hybrid and quark stars,''
	Phys.\ Rev.\ C {\bf 72} (2005) 025804
	[astro-ph/0501254].
	
	\bibitem{refine3} E.~S.~Fraga, A.~Kurkela and A.~Vuorinen,
	%``Interacting quark matter equation of state for compact stars,''
	Astrophys.\ J.\  {\bf 781} (2014) no.2,  L25
	[arXiv:1311.5154 [nucl-th]].
	
	\bibitem{art} A.~Vuorinen,
	%``Quark Matter Equation of State from Perturbative QCD,''
	arXiv:1611.04557 [hep-ph].
	
	\bibitem{prototype} O.~Zubairi, A.~Romero and F.~Weber,
	%``Static solutions of Einstein's field equations for compact stellar objects,''
	J.\ Phys.\ Conf.\ Ser.\  {\bf 615} (2015) no.1,  012003.
	
	\bibitem{postnewtonian} M.~Capone and M.~L.~Ruggiero,
	%``Jumping from Metric f(R) to Scalar-Tensor Theories and the relations between their post-Newtonian Parameters,''
	Class.\ Quant.\ Grav.\  {\bf 27} (2010) 125006
	[arXiv:0910.0434 [gr-qc]].
	
	\bibitem{cassini} B.~Bertotti, L.~Iess and P.~Tortora,
	%``A test of general relativity using radio links with the Cassini spacecraft,''
	Nature {\bf 425} (2003) 374.
	
	\bibitem{bullet1} M.~Markevitch {\it et al.},
	%``Direct constraints on the dark matter self-interaction cross-section from the merging galaxy cluster 1E0657-56,''
	Astrophys.\ J.\  {\bf 606} (2004) 819
	[astro-ph/0309303].
	
	\bibitem{bullet2} A.~Robertson, R.~Massey and V.~Eke,
	%``What does the Bullet Cluster tell us about self-interacting dark matter?,''
	arXiv:1605.04307 [astro-ph.CO].
	
	\bibitem{review} B.~L.~Young,
	%``A survey of dark matter and related topics in cosmology,''
	Front.\ Phys.\ (Beijing) {\bf 12} (2017) no.2,  121201.
	
	\bibitem{GR1} P.~C.~Peters and J.~Mathews,
	%``Gravitational radiation from point masses in a Keplerian orbit,''
	Phys.\ Rev.\  {\bf 131} (1963) 435.
	
	\bibitem{GR2} P.~C.~Peters,
	%``Gravitational Radiation and the Motion of Two Point Masses,''
	Phys.\ Rev.\  {\bf 136} (1964) B1224.
	
	\bibitem{Laurentis1} M.~De Laurentis and S.~Capozziello,
	%``Quadrupolar gravitational radiation as a test-bed for f(R)-gravity,''
	Astropart.\ Phys.\  {\bf 35} (2011) 257
	[arXiv:1104.1942 [gr-qc]].
	
	\bibitem{Laurentis2} M.~De Laurentis and I.~De Martino,
	%``Testing f(R)-theories using the first time derivative of the orbital period of the binary pulsars,''
	Mon.\ Not.\ Roy.\ Astron.\ Soc.\  {\bf 431} (2014) 741
	[arXiv:1302.0220 [gr-qc]].
	
	\bibitem{Laurentis3} M.~De Laurentis and I.~De Martino,
	%``Probing the physical and mathematical structure of $f(R)$-gravity by PSR J0348 + 0432,''
	Int.\ J.\ Geom.\ Meth.\ Mod.\ Phys.\  {\bf 12} (2015) no.04,  1550040
	[arXiv:1310.0711 [gr-qc]].
	
	\bibitem{TOV1} J.~R.~Oppenheimer and G.~M.~Volkoff,
	%``On Massive neutron cores,''
	Phys.\ Rev.\  {\bf 55} (1939) 374.
	
	\bibitem{TOV2} Tolman R. C. 1939 Phys. Rev. 55 364-73.
	
	\bibitem{2fluid1} F. Sandin and P. Ciarcelluti, Astropart. Phys., 32, 278 (2009).
	
	\bibitem{2fluid2} P. Ciarcelluti and F. Sandin, Phys. Lett. B 695, 19 (2011).
	
	\bibitem{Wang:2012ts} L.~Wang and X.~F.~Han,
	%``130 GeV gamma-ray line and enhancement of $h\to\gamma\gamma$ in the Higgs triplet model plus a scalar dark matter,''
	Phys.\ Rev.\ D {\bf 87} (2013) no.1,  015015
	[arXiv:1209.0376 [hep-ph]].
	
	\bibitem{LAT} 
	\url{https://www-glast.stanford.edu/}
\end{thebibliography}
\end{document}